\documentclass[12pt]{article}
\usepackage{graphicx}
\graphicspath{{./figures/}{./}}
\usepackage[sort&compress]{natbib}

\newcommand{\eq}[1]{(\ref{#1})} 
\newcommand{\mbf}[1]{\mbox{\boldmath$#1$\unboldmath}} 
\newcommand{\D}{\mathrm{d}} 

\begin{document}

\def\spacingset#1{\renewcommand{\baselinestretch}%
{#1}\small\normalsize} \spacingset{1}


\title{\bf (Un)Conditional Sample Generation Based on Distribution Element Trees}
\author{Daniel W.\ Meyer\thanks{
  The author gratefully acknowledges financial support by ETH Z\"urich. Moreover, he is thankful to Timon R\"uesch for his help during the preparation-phase of the R implementation.}\hspace{.2cm}\\
  Institute of Fluid Dynamics, ETH Z\"urich}
\maketitle

\bigskip
\begin{abstract}
Recently, distribution element trees (DETs) were introduced as an accurate and computationally efficient method for density estimation. In this work, we demonstrate that the DET formulation promotes an easy and inexpensive way to generate random samples similar to a smooth bootstrap. These samples can be generated unconditionally, but also, without further complications, conditionally utilizing available information about certain probability-space components.
\end{abstract}

\noindent%
\unitlength\textwidth
\begin{figure}
\begin{picture}(1,0.01)
\put(0,1.4){\fbox{
\begin{minipage}[b]{0.98\textwidth}
The final form of this manuscript was published by Taylor \& Francis in the \textbf{Journal of Computational and Graphical Statistics} on 08/06/2018 and is available online at http://www.tandfonline.com/10.1080/10618600.2018.1482768
\end{minipage}}}
\end{picture}
\end{figure}
\vfill

\newpage
\spacingset{1.45} 

\section{Introduction}

In many statistical problems, resampling methods such as cross-validation---used for example for the evaluation of regression models, jacknife---applied for estimator bias reduction \citep{Leger:1992a,Chernick:2012a}, or bootstrap methods \citep{Efron:1979a} are applied. Bootstrapping is for example applied to determine---based on an available ensemble of $n$ independent and identically distributed samples $\mathbf{x}_1,\mathbf{x}_2,\ldots,\mathbf{x}_n$ that stem from an unknown $d$-dimensional probability density $p(\mathbf{x})$ with $\mathbf{x}\in\Omega$---the distribution of an estimator \citep{Hinkley:1988a}. The resulting distribution can be further used to evaluate standard errors and confidence intervals, or for hypothesis testing.

The basic concept behind the bootstrap is the use of simulated bootstrap ensembles $\mathbf{x}_1^*,\mathbf{x}_2^*,\ldots,\mathbf{x}_n^*$ that are, in the case of the standard bootstrap, generated from the empirical distribution function $p_n(\mathbf{x}) = \frac{1}{n}\sum_{j=1}^n \delta(\mathbf{x}-\mathbf{x}_j)$ with the Dirac function $\delta$. Alternatively, bootstrap ensembles may be obtained from a parametric model (parametric bootstrap) or from a smoothed density resulting typically from kernel density estimation (KDE) \citep{Hesterberg:2015a} (smoothed bootstrap). Based on the available ensemble and the simulated bootstrap ensembles, the properties of a statistical quantity, e.g., a parameter estimator, can be extracted. The standard and smoothed bootstraps enable this extraction by means of Monte Carlo simulation without further parametric assumptions about the underlying distribution~$p(\mathbf{x})$, which renders both bootstraps particularly useful. Under certain conditions, \citet{Silverman:1987a} have demonstrated that the smoothed bootstrap leads to parameter estimates at reduced mean square errors compared to the standard bootstrap. \citet{Hall:1989a} have shown that the smoothed bootstrap has an improved error convergence for increasing~$n$ over the standard bootstrap, when the statistical quantity of interest involves the probability density such as when estimating quantiles. Moreover, smoothed bootstrapping is advantageous when applied for optimal bandwidth selection in KDE \citep{Leger:1992a,Taylor:1989a}.

In addition to existing density estimation methods such as KDE \citep[e.g.,][]{Silverman:1998a,Scott:2015a,Botev:2010a} or mixture-based approaches \citep[e.g.,][]{Wang:2015a,Neal:2000a,Ferguson:1973a}, we recently introduced a new non-parametric density estimator that is based on distribution element trees (DETs) \citep{Meyer:2017a}. Here, the $d$-dimensional probability space~$\Omega$ is decomposed into $m$ disjoint cuboids $C_k$ with $\Omega = \bigcup_{k = 1}^m C_k$. The density inside cuboid $k$ is given by
\begin{equation}\label{eqDEPDF}
p_k(\mathbf{x}) = \left\{\begin{array}{ll}
\displaystyle\frac{n(C_k)}{n}\prod_{i = 1}^d p[x_i|\mbf{\theta}_i(C_k)] & \forall\;\mathbf{x}\in C_k, \\
0 & \mbox{otherwise,}
\end{array}\right.
\end{equation}
and can be viewed as a simplest building block or atom of the DET estimator
\begin{equation}\label{eqDETPDF}
\hat{p}(\mathbf{x}) = \sum_{k = 1}^m p_k(\mathbf{x})\;\forall\;\mathbf{x}\in\Omega
\end{equation}
of the unknown density $p(\mathbf{x})$. In cuboid density~\eq{eqDEPDF}, $n(C_k)$ counts the number of samples $\mathbf{x}_j\in C_k$ with $j\in\{1,2,\ldots,n\}$ and $p[x_i|\mbf{\theta}_i(C_k)]$ is a marginal polynomial density for component~$x_i$ of the $d$-dimensional probability space~$\Omega$ \citep[e.g.,][equations~(3) and~(4)]{Meyer:2017a}. The density parameter vector~$\mbf{\theta}_i(C_k)$ is estimated from the $n(C_k)$ samples in~$C_k$. The cuboids $C_k$ and their densities $p_k(\mathbf{x})$, that form a so-called distribution element (DE), are resulting from a tree-like subdivision process with cuboids (tree nodes) being divided into subcuboids (tree branches). Subdivisions are determined by a goodness-of-fit test that evaluates locally the compatibility of density~\eq{eqDEPDF} with the samples contained in~$C_k$. This subdivision criterion resolves termination issues of earlier tree-based methods \citep[e.g.,][]{Ram:2011a,Wong:2010a,Jiang:2016a} and leads to a highly adaptive estimator with a density that is either piecewise constant, piecewise linear, etc.\ depending on the DE order or more precisely the order of $p[x_i|\mbf{\theta}_i(C_k)]$. The performance of the DET estimator was evaluated in a series of test cases of different dimensionality involving Gaussian mixtures as well as other densities including discontinuous ones. The linear DET estimator was found to be more accurate and computationally efficient than existing tree-based approaches such as density trees \citep{Ram:2011a} and limited-lookahead optional P\'olya trees \citep{Wong:2010a,Jiang:2016a}. Moreover, compared to the highly-cited adaptive KDE of \citet{Botev:2010a}, that is based on the solution of a non-linear diffusion partial differential equation (PDE) and a fix-point iteration involving a linear diffusion PDE, the linear DET estimator displayed equal or superior mean integrated square error (MISE) convergence rates and scaled favorably in terms of computational costs for increasing~$n$.

In this work, we are making use of the structure of the DET estimator given by equations~\eq{eqDEPDF} and~\eq{eqDETPDF} to formulate a smooth bootstrap that allows for the simulation of (un)conditional samples. The underlying formulation is outlined in section~\ref{secFormulation} followed by exemplary simulations presented in section~\ref{secExperiments}.

\section{Formulation}\label{secFormulation}

In the smoothed bootstrap resulting from KDE, a sample is generated by (1) drawing a sample from the available ensemble and by (2) adding noise to it, as implied by the kernel located at the selected sample \citep[equation~(3.11)]{Efron:1979a}\citep[section~3.9.1]{Scott:2015a}. Given the DET estimator~\eq{eqDETPDF}, samples can be generated based on a similar two-step procedure, as is outlined in a first step for the unconditional case in the following section.

\subsection{Unconditional Generation}

Given DET estimator~\eq{eqDETPDF}, the probability mass for a sample to reside in cuboid~$C_k$ is given by
\begin{equation}\label{eqDEkProbUc}
\int_{C_k} p_k(\mathbf{x}) \D\mathbf{x} = \frac{n(C_k)}{n} \int_{C_k} \prod_{i = 1}^d p[x_i|\mbf{\theta}_i(C_k)] \D\mathbf{x} = \frac{n(C_k)}{n} \prod_{i = 1}^d \int_{x_{i,l}^k}^{x_{i,u}^k} p[x_i|\mbf{\theta}_i(C_k)] \D x_i = \frac{n(C_k)}{n},
\end{equation}
where, equation~\eq{eqDEPDF} was introduced in the first step,
\begin{displaymath}
\int_{C_k} \ldots \D\mathbf{x} = \int_{x_{1,l}^k}^{x_{1,u}^k}\int_{x_{2,l}^k}^{x_{2,u}^k}\cdots\int_{x_{n,l}^k}^{x_{n,u}^k}\ldots\D x_n\cdots\D x_2\D x_1,
\end{displaymath}
cuboid~$k$ is defined as $C_k = \prod_{i = 1}^d [x_{i,l}^k,x_{i,u}^k]$ with lower and upper bounds $x_{i,l}^k$ and $x_{i,u}^k$, respectively, and the last step in expression~\eq{eqDEkProbUc} is based on the normalization condition of the marginal densities, i.e., $\int_{x_{i,l}^k}^{x_{i,u}^k} p[x_i|\mbf{\theta}_i(C_k)] \D x_i = 1$.

Therefore, in order to generate a random sample $\mathbf{x}^*$ based on an existing DET estimator, firstly we randomly pick a DE $C_k$ according to the probability masses $n(C_k)/n$. Inside the selected~$C_k$, components~$x_i$ are statistically independent (see equation~\eq{eqDEPDF}). The $x_i^*$ are thus generated in a second step based on uniformly distributed random numbers $y_i^*\in[0,1]$ and the quantile functions of the marginal densities $p[x_i|\mbf{\theta}_i(C_k)]$ as $x_i^* = P^{(-1)}[y_i^*|\mbf{\theta}_i(C_k)]$. For the constant and linear marginal densities considered in our earlier work \citep{Meyer:2017a}, the quantile functions are available in analytical form.

\subsection{Conditional Generation}

In a next step, a sample $\mathbf{x}^*$ shall be generated under the condition that components $x_{q+1},x_{q+2},\ldots,x_d$ with $1 \le q \le d$ take known prescribed values. Here, $q = d$ corresponds to the unconditional case discussed in the previous section, whereas with $q = 1$ all but one, or more precisely, all but the first component of $\mathbf{x}^*$ are known. It is pointed out that the ordering of components~$x_i$ with known components following unknown components serves the sole purpose of a simplified notation. The conditional density of the remaining random components $x_1,x_2,\ldots,x_q$ is then given by
\begin{equation}\label{eqCondPDF}
p(\mathbf{x}^\prime|\mathbf{x}^c) = p(\underbrace{x_1,x_2,\ldots,x_q}_{\displaystyle = \mathbf{x}^\prime}|\underbrace{x_{q+1},x_{q+2},\ldots,x_d}_{\displaystyle = \mathbf{x}^c}) = \frac{p(x_1,x_2,\ldots,x_d)}{p(x_{q+1},x_{q+2},\ldots,x_d)} = \frac{p(\mathbf{x})}{p(\mathbf{x}^c)},
\end{equation}
where
\begin{displaymath}
\mathbf{x} = \left(\begin{array}{c}\mathbf{x}^\prime\\ \mathbf{x}^c\end{array}\right) \mbox{, } p(\mathbf{x}^c) = \int_{x_{1,l}}^{x_{1,u}}\int_{x_{2,l}}^{x_{2,u}}\cdots\int_{x_{q,l}}^{x_{q,u}} p(\mathbf{x}) \,\D x_q\cdots\D x_2\D x_1
\end{displaymath}
is a normalization constant, and $x_{i,l}$ and $x_{i,u}$ are the lower and upper bounds, respectively, of component~$x_i$ of the probability space~$\Omega$. Similarly  to derivation~\eq{eqDEkProbUc} in the unconditional case, the probability mass for a sample to reside in cuboid $C_k$---with $C_k$ honoring condition
\begin{equation}\label{eqCondDE}
x_i \in [x_{i,l}^k,x_{i,u}^k]\;\forall\;i\in\{q+1,q+2,\ldots,d\},
\end{equation}
which accounts for the known prescribed components---is given by
\begin{eqnarray}\label{eqDEkProbC}
\lefteqn{\int_{x_{1,l}^k}^{x_{1,u}^k}\int_{x_{2,l}^k}^{x_{2,u}^k}\cdots\int_{x_{q,l}^k}^{x_{q,u}^k} p(\mathbf{x}^\prime|\mathbf{x}^c) \,\D x_q\cdots\D x_2\D x_1} \nonumber \\
 & = & \frac{1}{p(\mathbf{x}^c)} \int_{x_{1,l}^k}^{x_{1,u}^k}\int_{x_{2,l}^k}^{x_{2,u}^k}\cdots\int_{x_{q,l}^k}^{x_{q,u}^k} p(\mathbf{x}) \,\D x_q\cdots\D x_2\D x_1 \nonumber \\
 & = & \frac{1}{p(\mathbf{x}^c)} \int_{x_{1,l}^k}^{x_{1,u}^k}\int_{x_{2,l}^k}^{x_{2,u}^k}\cdots\int_{x_{q,l}^k}^{x_{q,u}^k} \frac{n(C_k)}{n}\prod_{i = 1}^d p[x_i|\mbf{\theta}_i(C_k)] \,\D x_q\cdots\D x_2\D x_1 \nonumber \\
 & = & \frac{1}{p(\mathbf{x}^c)} \frac{n(C_k)}{n} \prod_{i = q+1}^d p[x_i|\mbf{\theta}_i(C_k)] \prod_{i = 1}^q \int_{x_{i,l}^k}^{x_{i,u}^k} p[x_i|\mbf{\theta}_i(C_k)] \,\D x_i \nonumber \\
 & = & \frac{1}{p(\mathbf{x}^c)} \frac{n(C_k)}{n} \prod_{i = q+1}^d p[x_i|\mbf{\theta}_i(C_k)].
\end{eqnarray}
This result is based on equations~\eq{eqDEPDF}, \eq{eqDETPDF}, and~\eq{eqCondPDF}. Therefore in the conditional case, the probability masses from the unconditional case $n(C_k)/n$ are modified by the constant $1/p(\mathbf{x}^c)$ and a factor. The latter involves the marginal DE densities evaluated at the prescribed values and thus differs among the DEs that intersect with the conditions $\mathbf{x}^c$ or more precisely satisfy condition~\eq{eqCondDE}.

In summary, to generate a sample~$\mathbf{x}^*$ with components $x_{q+1},x_{q+2},\ldots,x_d$ prescribed, the cuboids or DEs satisfying condition~\eq{eqCondDE} are identified and one DE from this set is randomly picked according to the probabilities~\eq{eqDEkProbC}. Based on the marginal densities of the selected DE, components $x_1,x_2,\ldots,x_q$ of~$\mathbf{x}^*$ are determined in a second step based on uniformly-distributed random numbers $y_1^*,y_2^*,\ldots,y_m^*$ and quantile functions like in the unconditional case (see last paragraph of the previous section). Given the tree-based arrangement of the DEs introduced in our earlier work \citep[section~2.2]{Meyer:2017a}, the DEs that honor condition~\eq{eqCondDE} can be identified efficiently. To this end, we start at the tree root (cuboid corresponding to $\Omega$) and navigate along branches keeping track of cuboids that satisfy condition~\eq{eqCondDE}.

\section{Simulations}\label{secExperiments}

\begin{figure}
\unitlength\textwidth
\begin{center}
\includegraphics[width=0.6\textwidth]{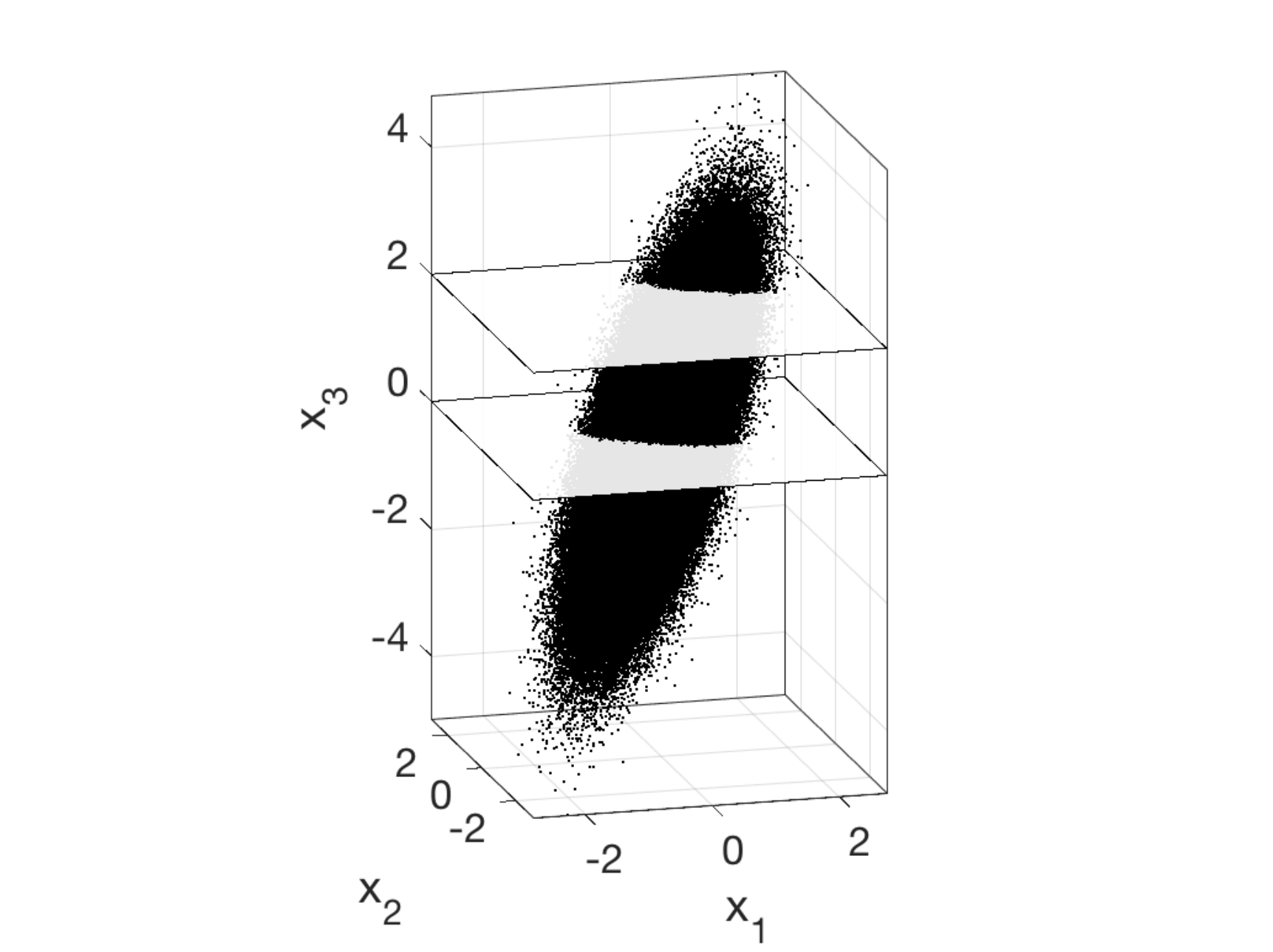}
\end{center}
\caption{The ensemble with $10^6$ samples resulting from the Gaussian density~\eq{eqC1PDF} is shown in $x_1$-$x_2$-$x_3$ probability space. Conditioning $x_1$-$x_2$-planes at $x_3 = 0$ and~2 are provided as well.}\label{figC1Data}
\end{figure}
\begin{figure}
\unitlength\textwidth
\begin{center}
\includegraphics[width=\textwidth]{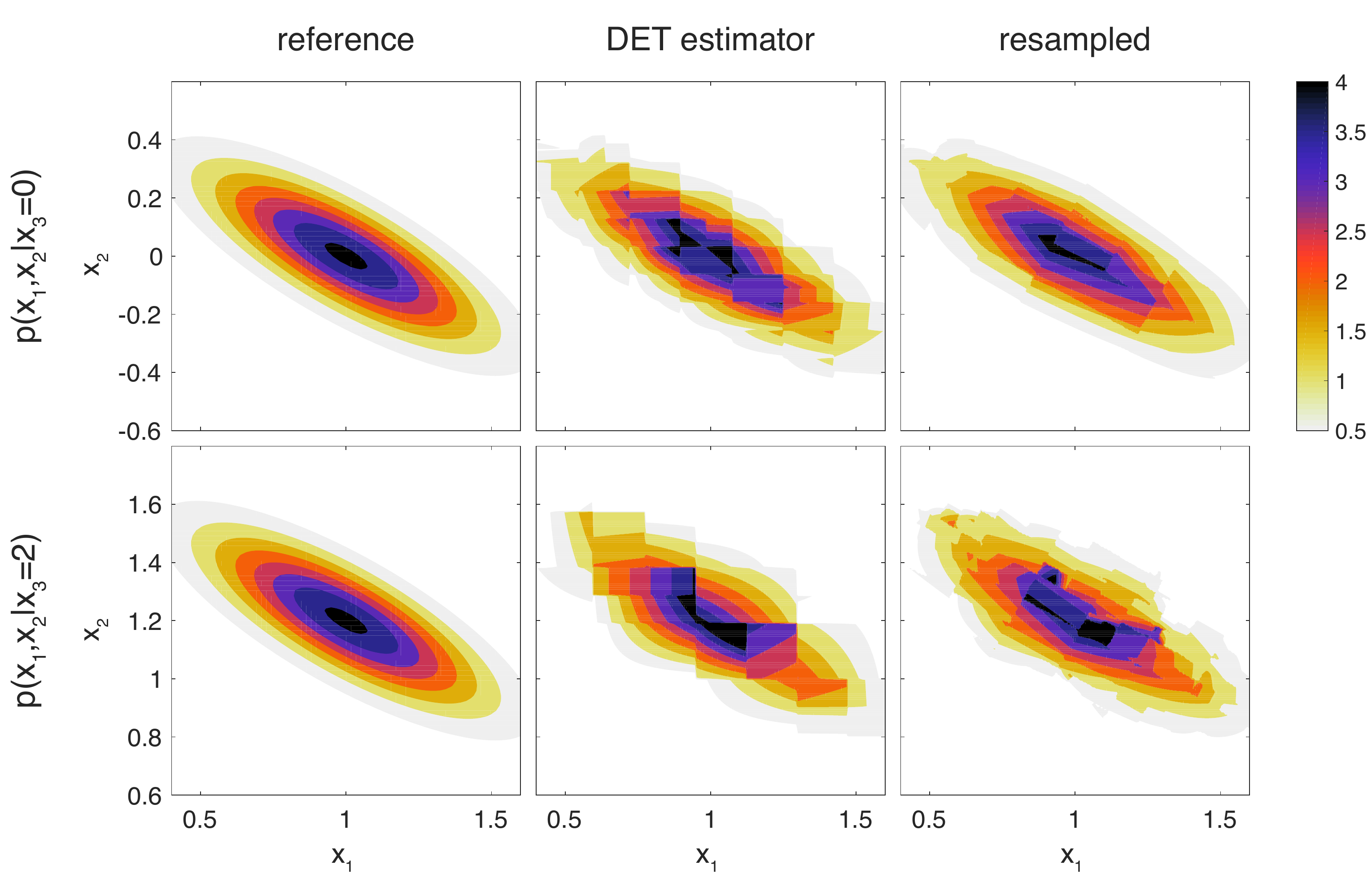}
\end{center}
\caption{Conditional density and density estimates of $p(x_1,x_2|x_3)$ resulting from the joint Gaussian~\eq{eqC1PDF} (left column), the DET estimator (center) resulting from the ensemble with $10^6$ samples $(x_1^*,x_2^*,x_3^*)^\top$, and the DET estimator (right) based on $10^5$ resampled $(x_1^*,x_2^*)^\top$ conditional on $x_3$ are shown. Densities and density estimates for $x_3 = 0$ and~2 are depicted in the top and bottom rows, respectively.}\label{figC1PDFs}
\end{figure}
As an illustration of the outlined (conditional) DET-based sample generation method, we provide two different examples involving linear DEs.

The first case deals with an available ensemble with $n = 10^6$ samples that stems from the Gaussian density
\begin{equation}\label{eqC1PDF}
p(\mathbf{x}) = \frac{\exp\left[-{\textstyle\frac{1}{2}}(\mathbf{x}-\mbf{\mu})^\top \mathbf{C}^{-1} (\mathbf{x}-\mbf{\mu})\right]}{\sqrt{(2\pi)^d \det(\mathbf{C})}}
\end{equation}
with $d = 3$, $\mathbf{x} = (x_1,x_2,x_3)^\top$, mean vector $\mbf{\mu} = (0,0,0)^\top$, and arbitrarily chosen covariance matrix
\begin{displaymath}
\mathbf{C} = \left(\begin{array}{ccc}0.35 & 0.25 & 0.5 \\ 0.25 & 0.4 & 0.6 \\ 0.5 & 0.6 & 1\end{array}\right).
\end{displaymath}
The degree of statistical dependence among the components~$x_i$ is illustrated in \figurename{}~\ref{figC1Data}, where the available ensemble is depicted. Given the data and using the DET implementation \citep{Meyer:2017c}, a linear DET was constructed with equal-size cuboid splits \citep[for details see][section~2.2.1]{Meyer:2017a}. Cross-sections at $x_3 = 0$ and~2 of the resulting density estimate are compared in \figurename{}~\ref{figC1PDFs} against the reference density~\eq{eqC1PDF}. Based on the DET and using the previously outlined methodology, $10^5$ samples~$\mathbf{x}^*$ were resampled under the condition that $x_3 = 0$ or~2. The resulting conditional densities $p(x_1^*,x_2^*|x_3)$, estimated again with a linear DET, are depicted in \figurename{}~\ref{figC1PDFs}. One can observe that at $x_3 = 0$, where the marginal density $p(x_3)$ is larger (like the local sample frequency) compared to $p(x_3 = 2)$, the statistical error is locally smaller in the DET estimator used for resampling, and consequently the agreement between the reference density (first column in \figurename{}~\ref{figC1PDFs}) and the resulting resampled data (last column) is better.

\begin{figure}
\unitlength\textwidth
\begin{picture}(1,0.36)
\put(0.0,0.0){\includegraphics[width=0.5\textwidth]{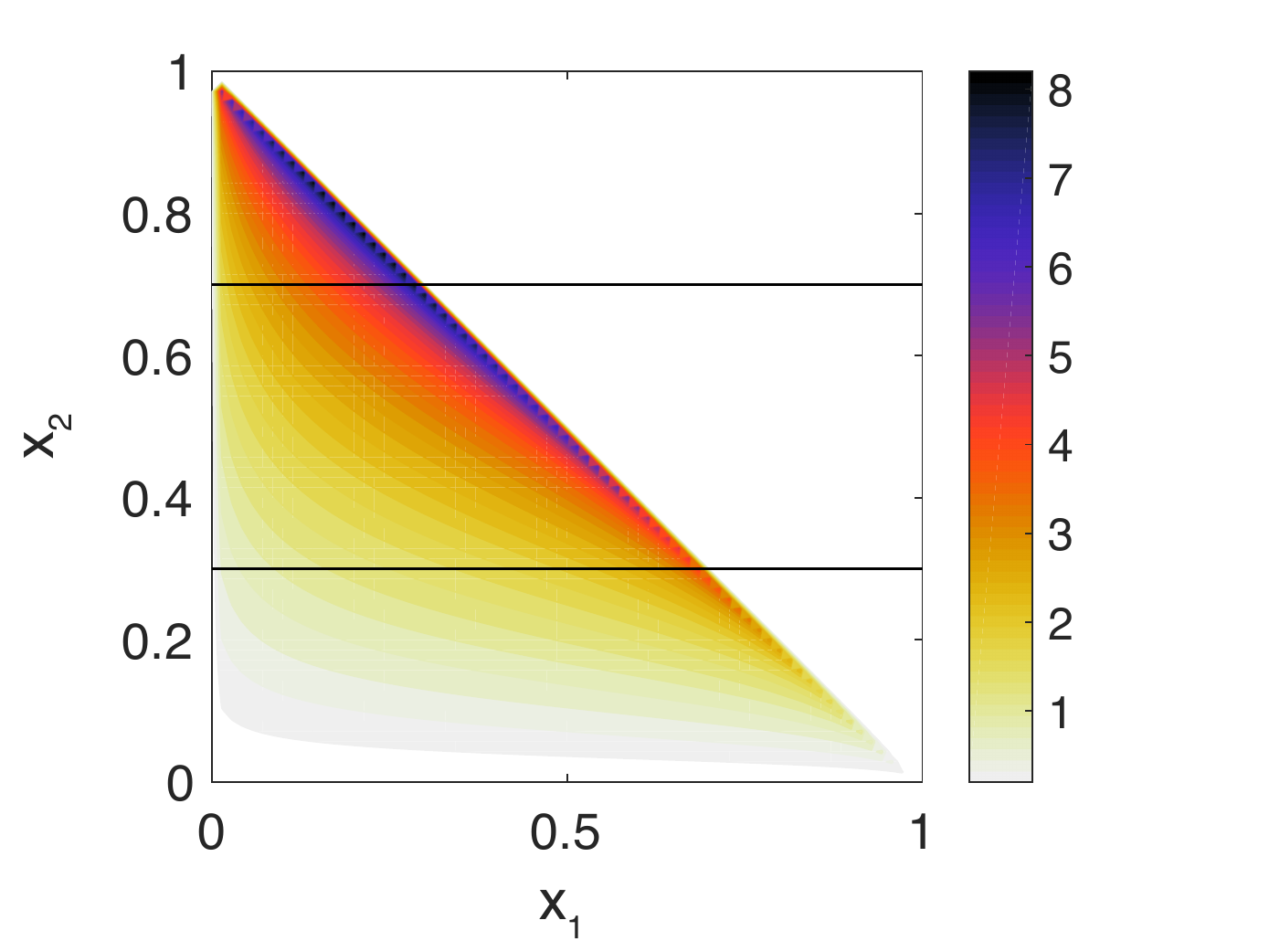}}
\put(0.01,0.35){\makebox(0,0){(a)}}
\put(0.56,0.0){\includegraphics[width=0.5\textwidth]{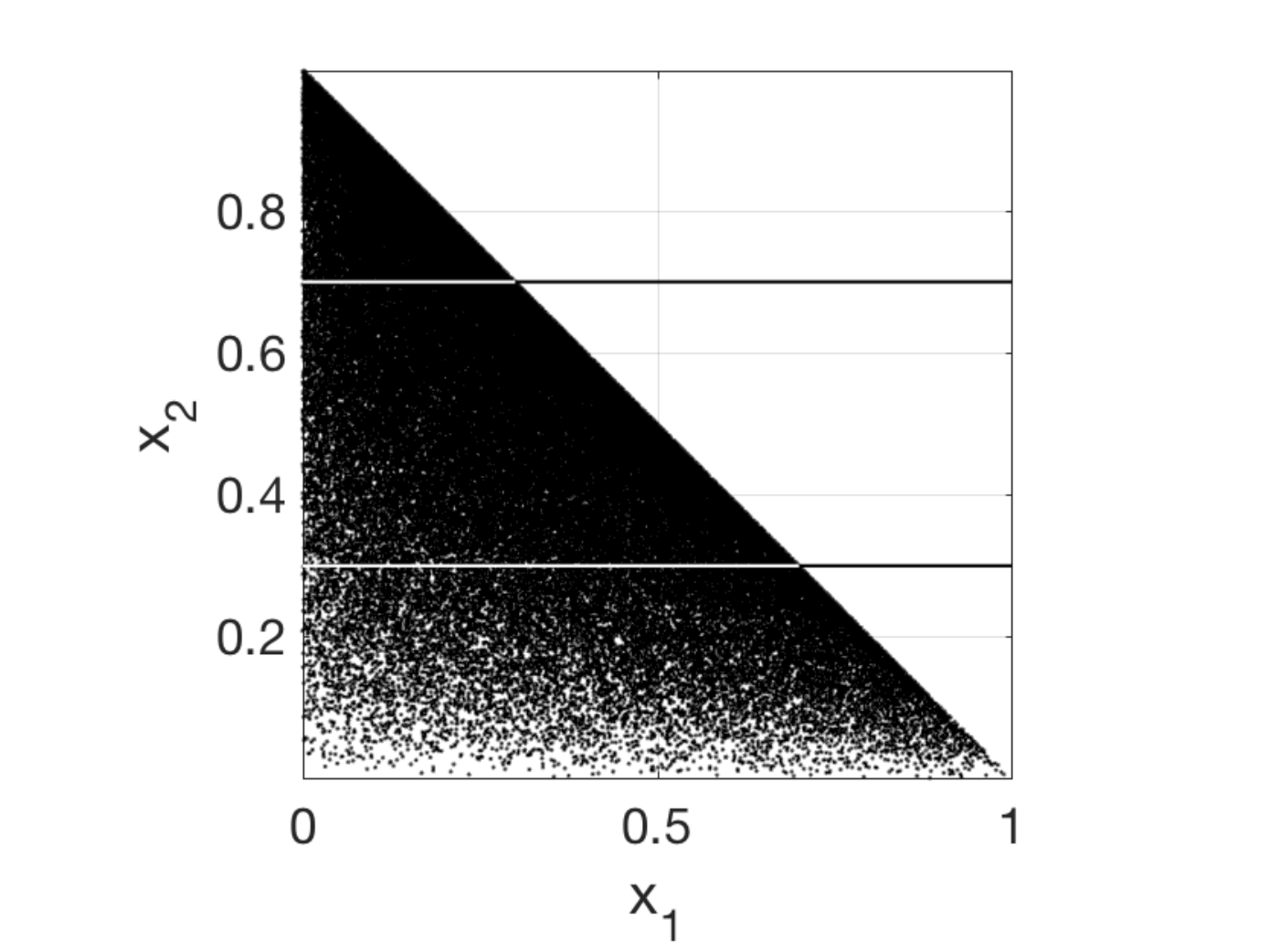}}
\put(0.57,0.35){\makebox(0,0){(b)}}
\end{picture}
\caption{(a) Bi-variate Dirichlet density~\eq{eqC2PDF} with parameters $\alpha_1 = 1.25$, $\alpha_2 = 2$, and $\alpha_3 = 0.75$, and (b) the corresponding ensemble with $n = 10^5$ samples. The horizontal lines mark conditioning locations at $x_2 = 0.3$ and~0.7.}\label{figC2PDF}
\end{figure}
\begin{figure}
\unitlength\textwidth
\begin{center}
\includegraphics[width=0.55\textwidth]{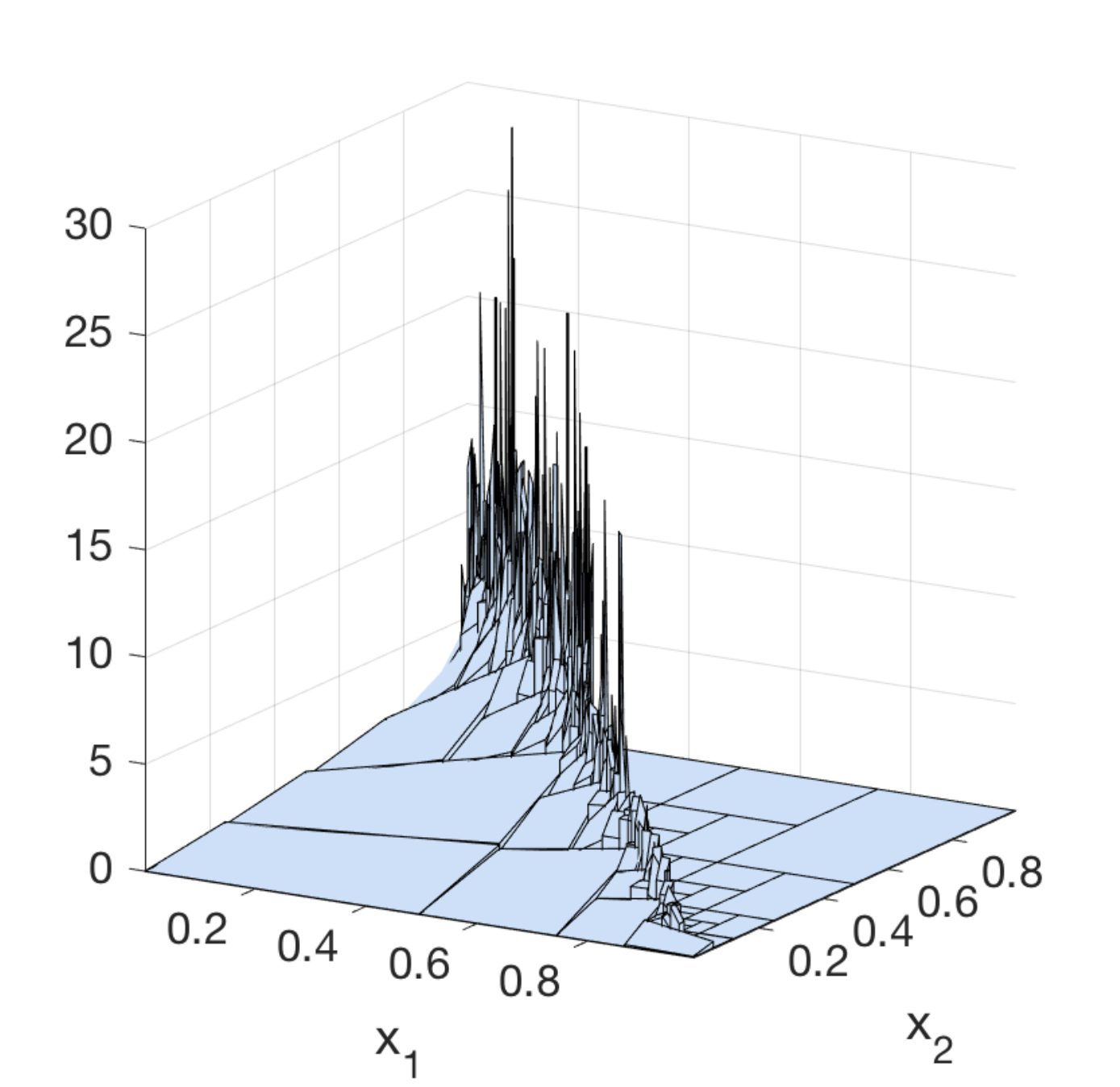}
\end{center}
\caption{Linear DET estimator resulting from the Dirichlet ensemble show in \figurename{}~\ref{figC2PDF}(b).}\label{figC2Det}
\end{figure}
\begin{figure}
\unitlength\textwidth
\begin{picture}(1,0.36)
\put(0.0,0.0){\includegraphics[width=0.5\textwidth]{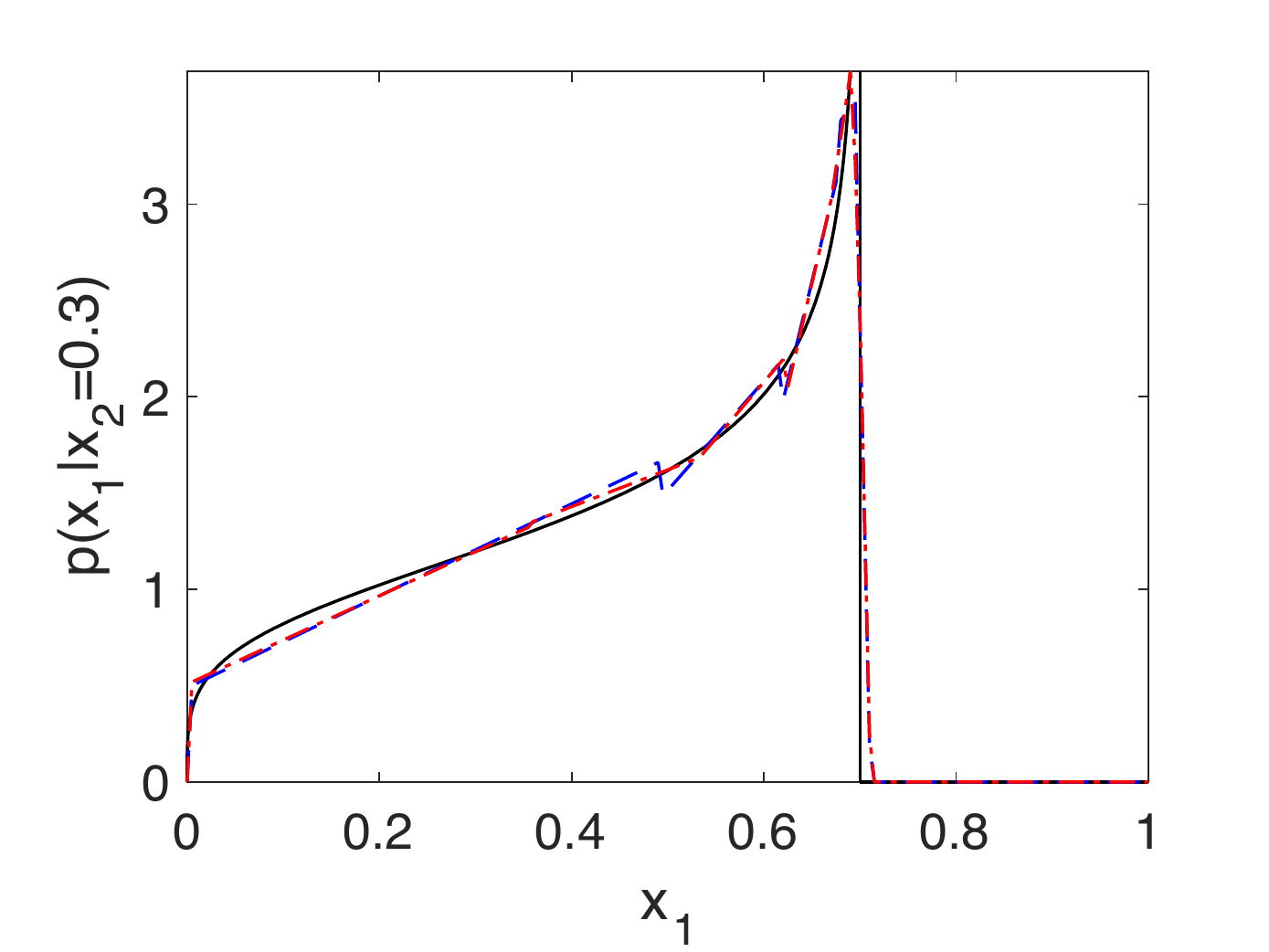}}
\put(0.01,0.35){\makebox(0,0){(a)}}
\put(0.56,0.0){\includegraphics[width=0.5\textwidth]{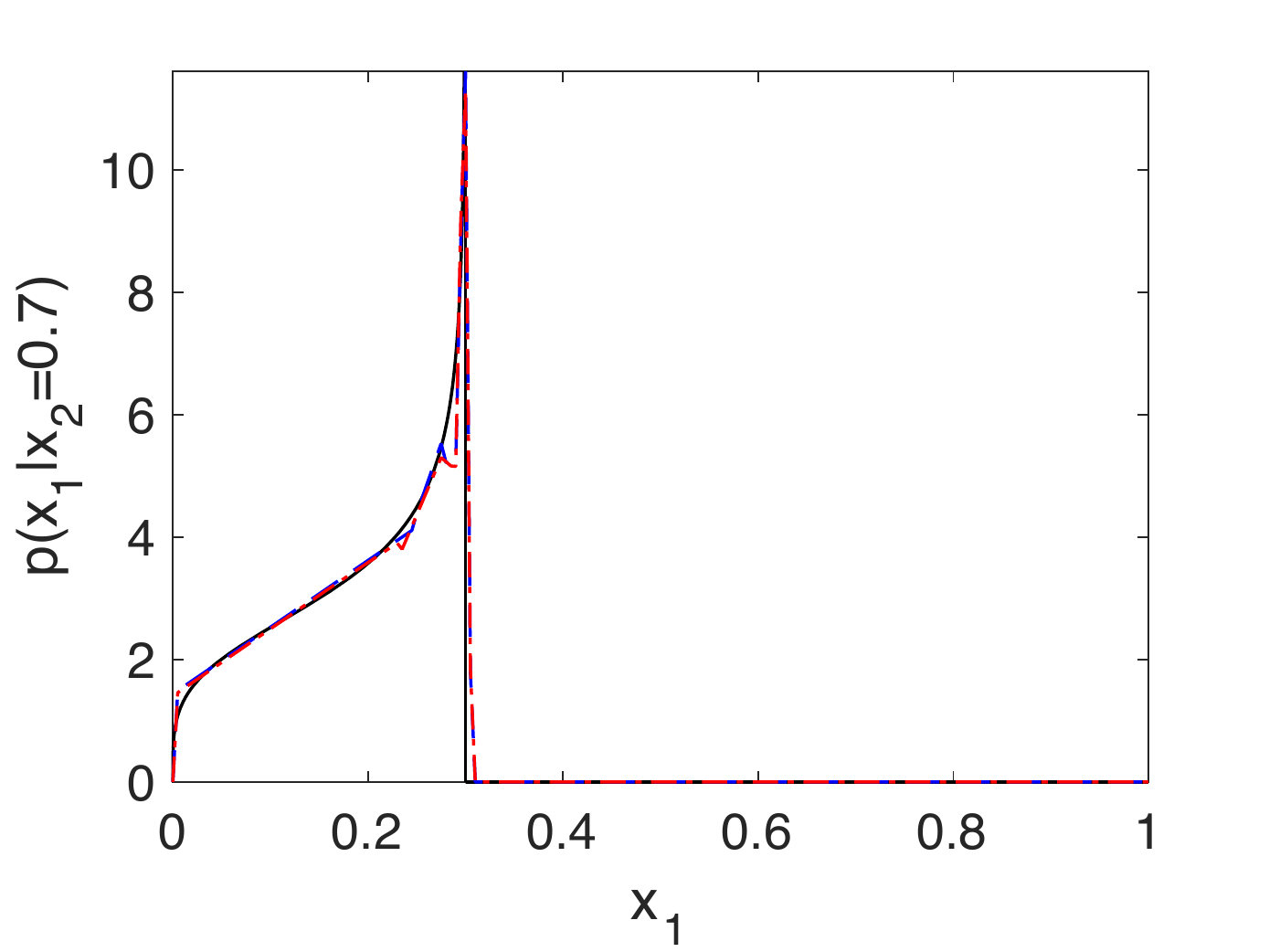}}
\put(0.57,0.35){\makebox(0,0){(b)}}
\end{picture}
\caption{Conditional density and density estimates of $p(x_1|x_2)$ resulting from the Dirichlet density~\eq{eqC2PDF} (black solid), the DET estimator (blue dashed) resulting from an ensemble with $10^5$ samples $(x_1^*,x_2^*)^\top$, and the DET estimator (red dash-dotted) based on $10^5$ resampled $x_1^*$ conditional on $x_2$ are shown. Densities and density estimates for $x_2 = 0.3$ and~0.7 are depicted in panels~(a) and~(b), respectively.}\label{figC2PDFs}
\end{figure}
In a second case, the bi-variate Dirichlet density
\begin{equation}\label{eqC2PDF}
p(x_1,x_2) = \left\{\begin{array}{ll}
\displaystyle\frac{x_1^{\alpha_1-1} x_2^{\alpha_2-1} (1-x_1-x_2)^{\alpha_3-1} \Gamma(\alpha_1+\alpha_2+\alpha_3)}{\Gamma(\alpha_1)\Gamma(\alpha_2)\Gamma(\alpha_3)} & \begin{array}{ll}\forall & x_1 \ge 0 \vee x_2 \ge 0 \\ & \vee\; 1-x_1-x_2 \ge 0,\end{array} \\
0 & \mbox{otherwise,}
\end{array}\right.
\end{equation}
with parameters $\alpha_1 = 1.25$, $\alpha_2 = 2$, and $\alpha_3 = 0.75$ is considered. This parameter set leads, differently to the previous Gaussian case, to a discontinuity along the line $x_2 = 1-x_1$ as is depicted in \figurename{}~\ref{figC2PDF}(a), where density~\eq{eqC2PDF} is plotted. To further illustrate the resampling method outlined in section~\ref{secFormulation}, an ensemble with $n = 10^5$ samples stemming from the Dirichlet density~\eq{eqC2PDF} and shown in \figurename{}~\ref{figC2PDF}(b) was used as a basis, and the linear DET estimator depicted in \figurename{}~\ref{figC2Det} was constructed. Estimates of the conditional density $p(x_1|x_2)$ for $x_2 = 0.3$ and~0.7 resulting from the DET are plotted in \figurename{}~\ref{figC2PDFs} (blue dashed lines) and can be compared with the conditional densities derived from reference density~\eq{eqC2PDF}. Based on the DET estimator, ensembles with $10^5$ samples $x_1^*$ were resampled given the condition $x_2 = 0.3$ or~0.7. DET-based density estimates from the resampled data are provided in \figurename{}~\ref{figC2PDFs} as well (red dash-dotted lines) and compare well against the reference density~\eq{eqC2PDF} (black lines). A closer look at the DET estimators, that is the one from the available ensemble $\mathbf{x} = (x_1,x_2)^\top$ (blue dashed) and the DET from the conditionally resampled $x_1^*$ data (red dash-dotted), illustrates that the resampled data emulates as expected the underlying DET given by the blue dashed lines. 

The presented simulations, including DET generation and resampling, were performed with the Matlab implementation that is publicly available on the MathWorks File Exchange \citep{Meyer:2017c}. A corresponding implementation for the open source statistics software R is available as well.

\bibliographystyle{agsm}

\bibliography{dwm_det}
\end{document}